# Universal frequency-dependent ac conductivity
# of conducting polymer networks

A. N. Papathanassiou [a)], I. Sakellis and J. Grammatikakis

University of Athens, Department of Physics, Section of Solid State Physics, Panepistimiopolis, GR 15784 Zografos, Athens, Greece

A model based on the aspect of the distribution of the length of conduction paths accessible for electric charge flow reproduces the universal power-law dispersive ac conductivity observed in polymer networks and, generally, in disordered matter. Power exponents larger than unity observed in some cases are physically acceptable within this model. A saturation high frequency region is also predicted, in agreement with experimental results. There does *not* exist a 'universal fractional power law' (and is useless searching for a unique common critical exponent), but a *qualitative* universal behavior of the ac condductivity in disordered media.

[a)] Corresponding author; e-mail address: antpapa@phys.uoa.gr

The measured ac conductivity $\sigma(\omega)$ of conducting and semi-conducting polymers is characterized by the transition above a critical (angular) frequency $\omega_0$ from a low-frequency dc plateau to a dispersive high-frequency region. As an example, the real part of the electrical conductivity $\sigma'$ is plotted as a function of $\omega$ for some conducting polymers and blends [1-3] is depicted in Fig. 1. The empirical Jonscher's universal law [4, 5] $\sigma_{ac} \propto \omega^n$, where $n$ is a fractional exponent roughly treated as constant less than 1, is often used to describe the ac component contributing to the dispersive region. Such behavior is observed in entirely different types of materials, such as disordered semiconductors, polymers, conducting polymer compounds ceramics, ion conducting glasses, heavily doped ionic crystals etc. [6-8], indicating that the *qualitative* characteristics of the universal response is irrelevant to the constituting atomic units. It merely has to do with the morphology of the conduction network. Although many different theoretical approaches [9-14] tried to conclude in a unique fractional exponent value (around 0.7) and justify the empirical universal law of Jonscher, there are *serious inefficiencies* about the validity of the universal power law:

(a) $n$ can hold values larger than unity (and there is *no physical argument* to restrict the value of $n$ below 1), e.g. in glassy $0.3(xLi_2O.(1-x)Li_2O)$ $0.7B_2O_3$ [15], in mixed compounds of $(NH_4)_3H(SO_4)_{1.42}(SeO_4)_{0.58}$ [16] and $K_3H(SeO_4)_2$ single crystals [17]

(b) $n$ is frequency dependent and

(c) what is the upper frequency limit of the 'universal' power law applies. For example, in Fig. 1 there is evidence at room temperature for a saturation high-frequency region in polypyrrole, which is more clear at low temperature (82K).

In this letter, stimulated by the topology of polymer chain morphology, we work on an ideal network of *conduction paths of various lengths*, which are accessible to electric charge carriers. Such picture may be representative in any disordered material. The simulation does not only reproduce the dispersive ac conductivity region but explains the above critical points, which remained still obscure.

The problem is focused on modeling the electric charge within a frame of accessible conduction paths flow under the influence of an external ac field. The length of the conduction paths is determined by spatial and energy distribution of the potential energy profile. For example, in a polymer network, different lengths of conduction paths are available due to possible distribution of the length of polymeric chains, cross-linking, mechanical bonding between different chains etc. In glasses or amorphous semiconductors the distribution of accessible conduction paths stems from the absence of periodicity in space and the presence of defects. Regardless the identity of the structural units and the way that they built up the solid and the type of electric charge entities that respond to the external field, we can adopt a unified manner to visualize how electric charge entities move in a network consisting of conduction paths of variable lengths. As we are going to see, a distribution in the path length reproduces the low-frequency plateau and the power law increases with frequency-dependent fractional exponent and - the most important - predicts the high frequency saturation.

A *qualitative* description of a polymer network structure is that of a group of polymer chains of various lengths, with conformational disorder and random orientation. The density of charge carriers (per unit length) is treated as constant. An electric charge carrier (such as a polaron or bipolaron) can hop along each chain (intra-chain transfer) and over cross-linked chain clusters. Inter-chain conduction is controlled by the degree of chain coupling. *Quantitative* differences arise from the different sample preparation conditions and the doping procedure. Electric charge flows along a network formed of conductive paths of different lengths L, which follow some distribution. A path does not necessarily coincide with an individual chain, but can probably be a cluster of coupled chains. Depending on its length, conformational disorder and orientation, a path can be long enough to connect the opposite sides of the specimen. Some shorter (in comparison with the specimen's dimensions) paths have dead ends. Moreover, we assume that the spatial distribution of the potential energy is the same regardless the length of the path.

It is an experimental fact that the ac conductivity of conducting polymers (more generally, of many disordered solids) as a function of frequency consists of a frequency independent low-frequency region and, above a critical frequency value $\omega_c$, a high-frequency one, which is dispersive. For a given angular frequency $\omega \prec \omega_c$, the measured conductivity $\sigma(\omega)$ results from the macroscopic conductivity (along paths connecting the opposite surfaces of the specimen, where electrodes are attached) and from the charge flow along paths, which are larger than $\upsilon/\omega$, where $\upsilon$ is some typical value for the mean velocity of the transferring charge carriers. In the dispesive region ($\omega > \omega_0$), the measured ac conductivity is the sum of the macroscopic conductivity and the conductivity along paths with length equal or larger than $\upsilon/\omega$ (i.e., the lengths corresponding to frequencies from $\omega_0$ to $\omega$).

The *measured* conductance $G(\omega)$, which is a quantity measured directly in dielectric experiments, consists of two components: Paths extending along the volume of the specimen, which contribute to the macroscopic (dc) conductance $G_{dc}$, and paths of length $L_k$ equal or longer than a critical length $L_c = \upsilon/\omega_c$, which contribute to conductance $G_k \propto L_k^{-1}$.

$$G(\omega) = G_{dc} + \sum_{L_k \geq L_c} G_k(L_k) \tag{1}$$

Assuming that the length follows a logarithmic distribution function f(logL), eq. (1) is modified to:

$$G(\omega) = G_{dc} + \alpha \int_{L_{min}}^{L} L^{-1} f(\log L) dL = G_{dc} + \alpha \int_{\log L_{min}}^{\log L} \log f(\log L) d(\log L) \tag{2}$$

where $L_{min}$ is the minimum path length, which is of the order of inter-atomic spacing and *α* is a constant. To a first approximation, path lengths are assumed to obey the normal distribution $f(\log L) = \frac{1}{\sqrt{2\pi}\sigma} \exp\left(-(\log L - \log L_0)^2 / (2\sigma^2)\right)$ characterized by a mean value $\log L_0$ ($L_0$ corresponding to the length of the most abundant paths) and broadening parameter σ: By applying the relation $\upsilon = L\omega$, we get:

$$G(\omega) = G_{dc} + A \int_0^\omega \frac{1}{\sqrt{2\pi}\sigma} \exp\left(-\frac{(\log \omega - \log \omega_0)^2}{2\sigma^2}\right) d\omega \tag{3}$$

where $\omega_0 = \upsilon/L_0$, and *A* is a constant related with the conductivity of an individual path, its effective cross-section area when regarded as a nano-wire and the velocity a charge carrier moves along each path.

Paths shorter than $L_c$ (i.e., $L_k<L_c$) contribute to capacitive effects, giving rise to polarization phenomena expressed by the real part of the complex permittivity $\epsilon'(\omega) \propto C(\omega)$, where $C(\omega)$ denotes the capacitance. $C(\omega)$. The capacitance of an individual of these paths is inversely proportional to its length. So, the formulation of $C(\omega)$ resembles that of Eqs. (2) and (3), by integrating from $\omega$ to infinite:

$$C(\omega) = C_0 \int_{\omega}^{\infty} \frac{1}{\omega\sqrt{2\pi}\sigma} \exp\left(-\frac{(\log\omega - \log\omega_0)^2}{2\sigma^2}\right) d\omega \qquad (4)$$

where $C_0$ is a constant, related with the saturation polarization of the dead-end paths achieved in the dc limit.

The total conductance vs frequency plot produced by employing Eq. (3) is depicted in a log-log representation in Fig. 2, for $\log L_0 = -5$ and $\sigma = 1$. We observe that t the low frequency plateau, where dc conduction dominates, is reproduced. The integration over the angular frequency yields an increase of $G(\omega)$ more paths gradually contribute to the total conductance on increasing $\omega$. In the high frequency limit, where the shorter length paths (comparable to the inter-atomic separation) significantly participate to the measured conductance, saturation is observed. The inset of Fig. 2 shows that the slope $d\log G/d\log\omega$ is a function of frequency, with maximum 0.75. The simulation reproduces *qualitatively* the dependence of $G(\omega)$ in a polymer chain network. The structural properties of the polymer chains, the distribution of the path lengths, the conductivity of an individual path (which can be regarded as a nano-wire, the density of states and the mobility of the electric charge carriers are the specific characteristics of an individual polymer, and determine the quantitative behavior of $G(\omega)$. The interplay between the above mentioned parameters yields different values of the critical transition frequency $\omega_0$, variable saturation region and frequency dependent slopes $d\log G(\omega)/d\log\omega$, which can be even larger than unity, explaining the 'unusual' behavior some materials exhibit [15-17], is physically acceptable within the present model.

The imaginary part of the complex permittivity $\epsilon'' \propto G(\omega)/\omega$, when plotted against $\log\omega$ (Fig. 3), consists of a decreasing dc component (with slope equal to -1 and dielectric relaxation component appearing as a 'knee'. By simply subtracting the dc component from the measured conductance, we get a well-defined dielectric dispersion $(G(\omega)-G_{dc})/\omega$. What is hidden behind this relaxation is the maximum absorption of the energy of the external harmonic electric field by the charge carriers flowing along the paths with length characterized by the distribution function $f(\log L)$. The inset of Fig. 3 shows the drop of $C(\omega)$ (which is proportional to $\varepsilon''$) on increasing frequency, according to Eq. (4).

The advantage of the model presented in this Letter is the *unified description* of the ac response of polymer networks and, by generalizing the idea of conducting paths, to *any non-crystalline material*. The advantages in relation with the long-standing power-law models that were published till now [9-14] are the following:

(i) The slope of the ac conductivity $d\log G(\omega)/d\log\omega$ is a function of frequency. The maximum slope given by differentiating Eq. (3) approaches a typical value for the exponent *n*, when the rough universal power law is employed to analyze the data.

(ii) The power exponent *n* is *neither limited to values below 1, nor is accumulating to some critical value* (around 0.6-0.7). Depending on the interplay between the parameters, n>1 is predicted from Eq. (3). The experimental results reporting n>1 [15-17], find their theoretical justification now.

(iii) There is an upper frequency limit for the dispersive conductivity. In the high-frequency limit, the measured conductivity practically saturates. Thus, the simulation presented in this Letter explains the appearance of *saturation region* observed in some cases such as in conducting polypyrrole (Fig. 1) and polyaniline [18] (the capability of detecting the saturation high frequency region is a matter of whether instrumentation is broadband.).

**Figure Captions**

**Figure 1:** The real part of the electrical conductivity $\sigma'$ vs (angular) frequency measured at room temperature. (a): conducting polyaniline [1]; (b): 10 % wt zeolite - 90% wt conducting polypyrrole blend [2], (c): 20 % wt conducting polypyrrole - 80 % wt polyaniline blend [1]; (d), (e) and (f): conducting polypyrrole with various degree of doping [3]. The conductivity values corresponding to curves (a) and (b) were multiplied by 10.

**Figure 2:** The logarithm of the measured conductance $G(\omega)$ as a function of the (angular) frequency $\omega$, for a normal log-distribution of paths around $\log L_0 = -5$ and broadening parameter $\sigma = 1$. The derivative of the theoretical curve is depicted in the inset diagram.

**Figure 3:** A dielectric loss representation of the data appearing in Fig. 2. $G/\omega$ is proportional to the imaginary part of the complex permittivity $\varepsilon''$ (left vertical axis). A well-defined dielectric absorption is obtained when the dc component $G_{dc}$ is subtracted from the total conductance $G(\omega)$ (right vertical axis). The capacitance C (which is proportional to the real part of the permittivity $\varepsilon'$) is plotted in the inset.

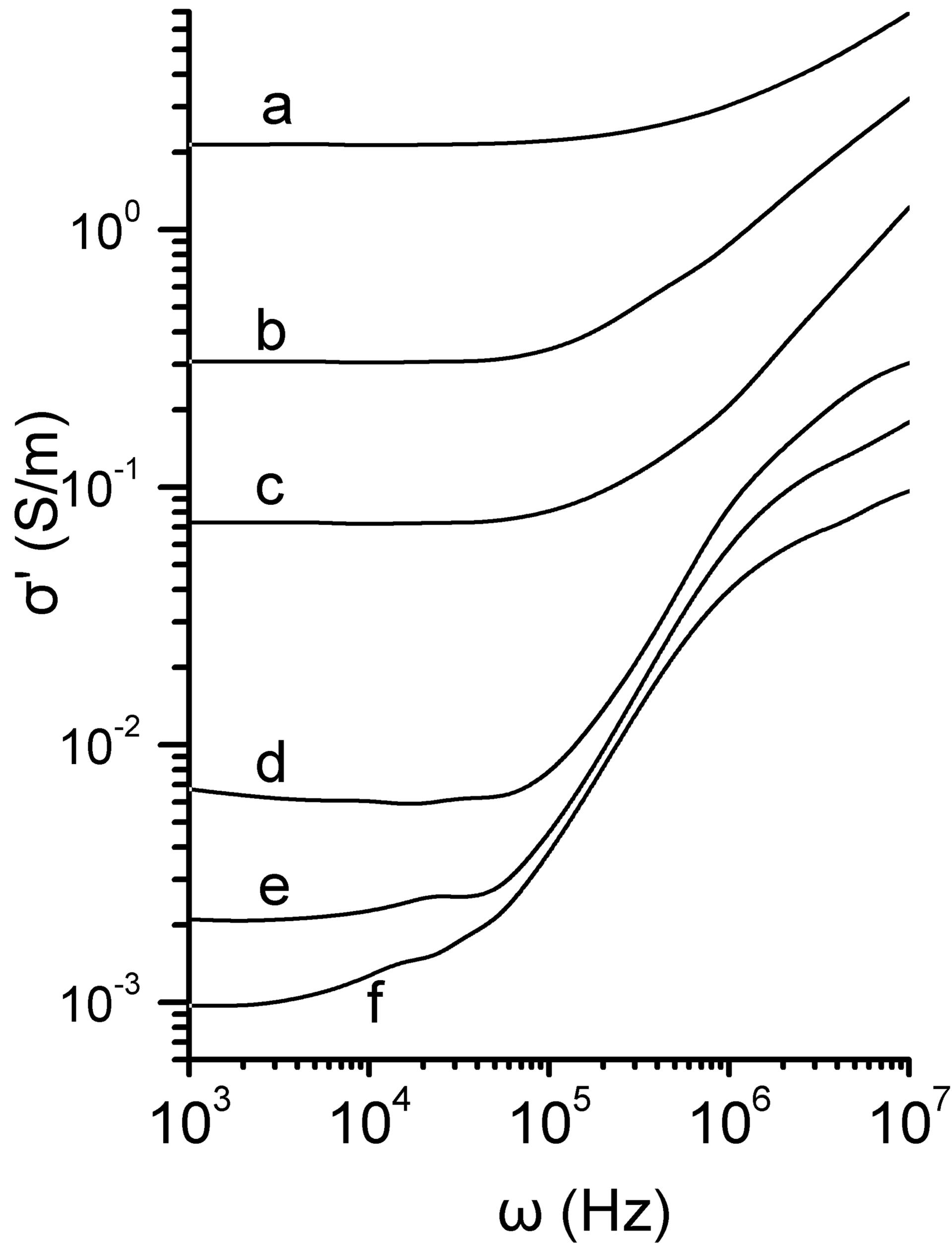

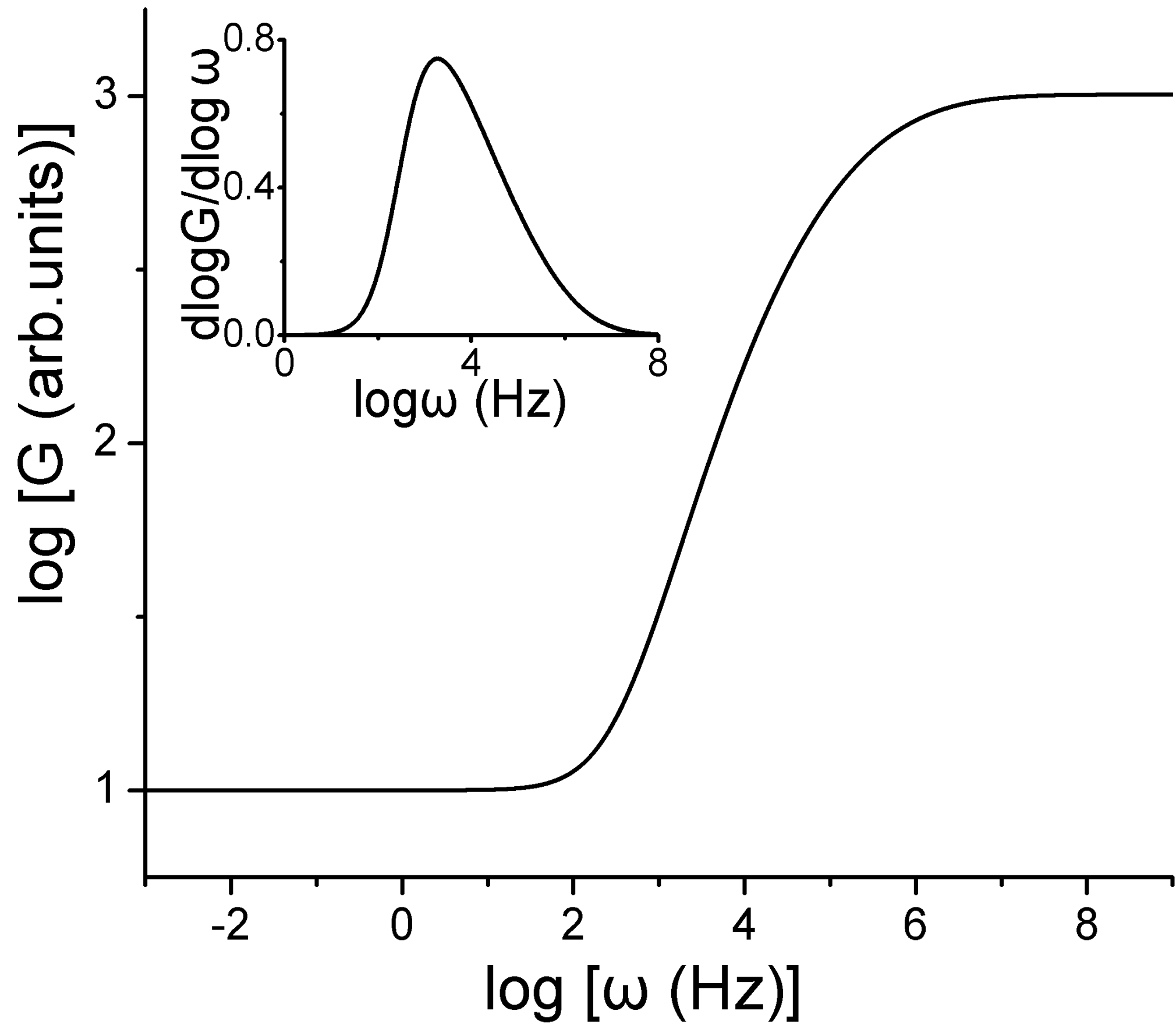

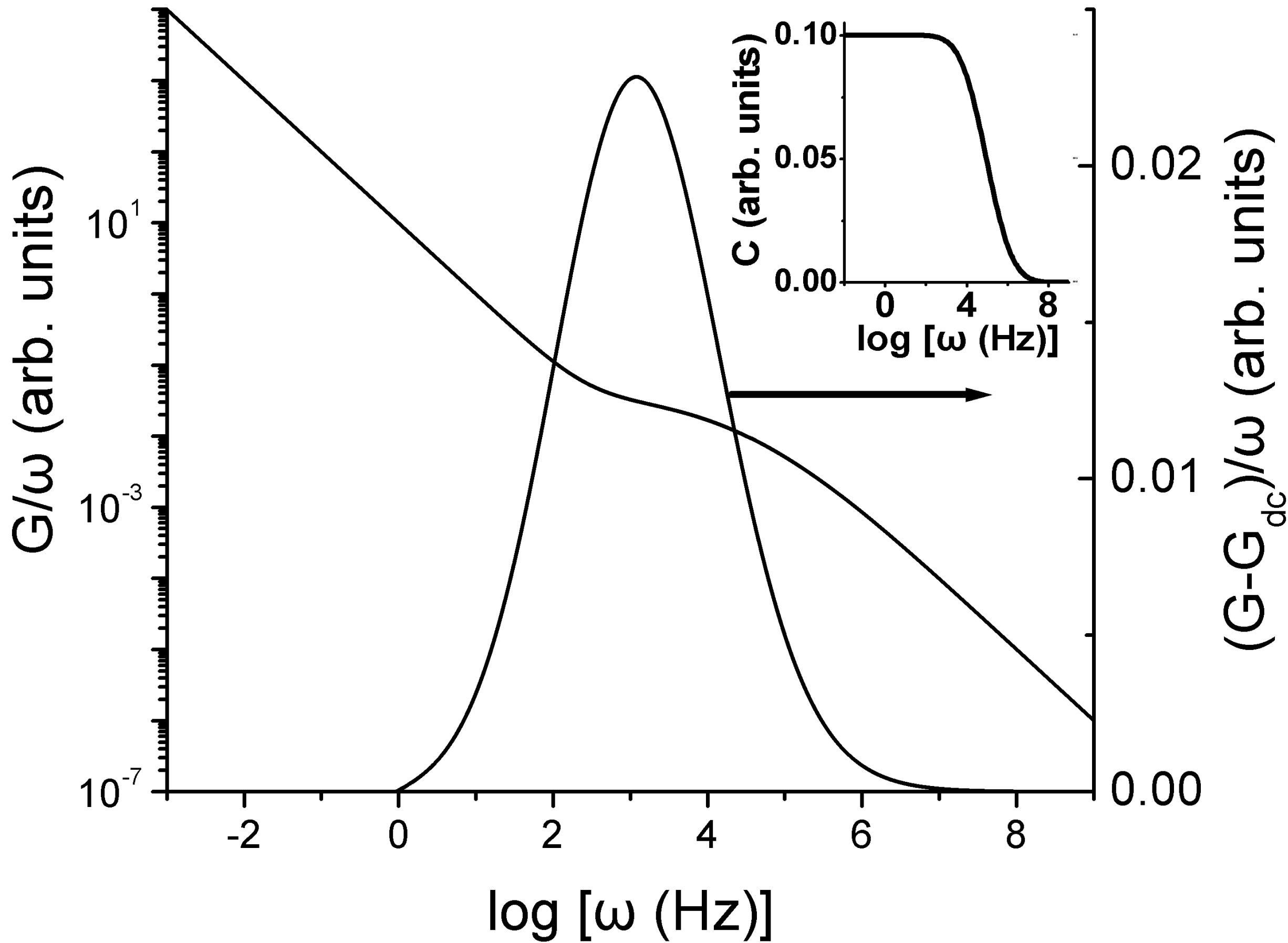